\documentclass[%
 reprint,
showpacs,preprintnumbers,
 amsmath,amssymb,
 aps,
]{revtex4-1}
\usepackage{xcolor}

\usepackage{graphicx}
\usepackage{dcolumn}
\usepackage{bm}

\newcommand\varpm{\mathbin{\vcenter{\hbox{%
  \oalign{\hfil$\scriptstyle+$\hfil\cr
          \noalign{\kern-.3ex}
          $\scriptscriptstyle({-})$\cr}%
}}}}
\newcommand\varmp{\mathbin{\vcenter{\hbox{%
  \oalign{$\scriptstyle({+})$\cr
          \noalign{\kern-.3ex}
          \hfil$\scriptscriptstyle-$\hfil\cr}%
}}}}
\usepackage[pagebackref=false,colorlinks,linkcolor=blue,filecolor=green,urlcolor=blue ,citecolor=blue]{hyperref}

\usepackage{changes}
\begin{document}

\title{Floquet topological phase transitions in a periodically quenched dimer}

\author{Milad Jangjan$^1$, Luis E. F. Foa Torres$^2$, and Mir Vahid Hosseini$^1$}
 \email[Corresponding author: ]{mv.hosseini@znu.ac.ir}
\affiliation{$^1$Department of Physics, Faculty of Science, University of Zanjan, Zanjan 45371-38791, Iran}
\affiliation{$^2$Departamento de F\'{\i}sica, Facultad de Ciencias F\'{\i}sicas y Matem\'aticas, Universidad de Chile, 837.0415 Santiago, Chile}

\begin{abstract}
We report on the theoretical investigation of the topological properties of a periodically quenched one-dimensional dimerized lattice where a piece-wise constant Hamiltonian switches from $h_1$ to $h_2$ at a partition time $t_p$ within each driving period $T$. We examine different dimerization patterns for $h_1$ and $h_2$ and the interplay with the driving parameters that lead to the emergence of topological states both at zero energy and at the edge of the Brillouin-Floquet quasi-energy zone. We illustrate different phenomena, including the occurrence of both edge states in a semimetal spectrum, the topological transitions, and the generation of zero-energy topological states from trivial snapshots. The role of the different symmetries in our results is also discussed.

\end{abstract}
\maketitle

\section {Introduction} \label{s1}

Topological states of matter in dynamically driven systems, known as Floquet topological states, have attracted much interest in recent years \cite{FloqReview1,FloqReview2}. A large number of approaches to engineer Floquet systems have been proposed theoretically \cite{FloState,FloqTheo0,FloqTheo1,FloqTheo2,FloqTheo3,FloqTheo5,FloqTheo6,FloqTheo7,FloqTheo8,FloqTheo9,FloqTheo10,FloqTheo11,FloqTheo12,FloqTheo13,FloqTheo14,FloqTheo15} and realized experimentally \cite{FloqExp1,FloqExp2,FloqExp3,FloqExp5,FloqExp6,FloqExp7,FloqExp8,FloqExp9}. Similar to static counterparts \cite{TopoSolid,StaticTI2}, in most of periodically driven systems, the so-called Floquet topological phases can emerge via the closing and reopening of the gap at topological phase transition points revealing topological edge states in the gap exhibiting exotic quantum Hall plateaus \cite{QuanHallPlat} and suppression of bulk transport at some Floquet topological transitions \cite{SuppTrans} even with richer bulk-boundary correspondence \cite{RichBBC}.

There are continuous \cite{FloqReview1,FloqReview2,FloState,FloqTheo1} and quenching \cite{QuanQuench1,QuanQuench2} protocols to drive systems periodically. In the latter class, the system experiences different static situations before and after a sudden quantum quench resulting in dynamical quantum phase transitions \cite{DQPT} obeying scaling and universality \cite{DQPT1,DQPT2}. Although the dynamical quantum phase transition can also take place under continuous driving \cite{ContinDQPT}, quantum quenched systems add many possibilities and is an active area of study. Indeed, a variety of interesting physics has been unveiled including a relation between dynamical microscopic probabilities and macroscopic properties \cite{Quench1}, first order dynamical phase transitions \cite{Quench2}, new topological phases such as the anomalous Floquet Anderson insulator~\cite{AFAI1} and its Hall response~\cite{AFAI2}, topological nodes \cite{Quench3}, non-Hermitian dynamical phase transitions \cite{Quench4}, and the effect of zero-energy mode \cite{Quench5}. Furthermore, large Chern numbers \cite{Quench6}, non-Hermitian Floquet topological phases \cite{PerQuench0,PerQuench,PerQuench1D,PerQuench1D2D3D}, and multiple Floquet dynamical quantum phase transitions \cite{Quench10} can take place for periodic quenching in one dimension.

On the other hand, one-dimensional (1D) platforms, despite their simplicity, can be employed to explore exotic phenomena \cite{exotic1D}. For instance, 1D dimerized lattice, known as Su-Schrieffer-Heeger (SSH) model \cite{SSH,SSH1} and its extensions \cite{GeSSH,GeSSH0,SSHZeeman,GeSSH1} can exhibit electric polarizations \cite{ElecPolar1} with quantized values \cite{ElecPolar2,ElecPolar3} in connection with nontrivial geometric phase of electronic bands \cite{Zak}. Furthermore, 1D systems under periodically applied perturbations \cite{Floq1D0,Floq1D1,Floq1D2,Floq1D21} including SSH model \cite{Floq1D3,Floq1D4} with chiral symmetry \cite{Floq1D0,Floq1D1,Floq1D2} have been studied revealing chiral flow~\cite{RichBBC}.

For Floquet dynamical systems having chiral symmetry \cite{Floq1D0}, there are two band gaps, around the center and boundary of the Floquet-Brillouin Zone (BZ). Under open boundary conditions, after closing and reopening the gaps, topological edge states with zero group velocity can be revealed inside the gaps at zero and $\pi$ quasi-energies in topologically nontrivial regimes \cite{Floq1D1}, whereas static systems can only give rise to zero-energy edge states. Correspondingly, these states can be characterized by two topological invariants, namely, the zero and $\pi$ winding numbers \cite{Floq1D2}, taking quantized integer values. Also, their classification scheme is either similar to or distinct from their static counterparts \cite{classification1,classification2}. However, it is interesting to have a control over turning on or off of the zero and $\pi$ modes separately such that the $\pi$ mode can also coexist with bulk states.

In this work, we consider Floquet topological phases of a periodically quenched 1D model. Within each period, the Hamiltonian switches from $h_1$ to $h_2$, where $h_1$ and $h_2$ are time-independent and the switching occurs after a time $t_p$ (which can also be used as a control parameter). In this case one could take $h_1$ and $h_2$ as dimers (e.g. SSH model). If $h_1=h_2$ then the Hamiltonian is effectively time-independent and gives the known SSH model. As the $h_1$ starts to differ from $h_2$, one gets zero energy edge states embedded in a metal, a semimetal, or a gapped system, depending on the frequency. A more interesting outcome occurs when $h_1$ and $h_2$ have opposite dimerizations. In this case, we find that the interplay between the driving period $T$ and the partition time $t_p$ leads to a rich diagram of edge states and topological transitions, both at zero energy and at the edge of the Floquet-BZ (the so called $\pi$-modes). We also show a case where topological states are generated from the driving protocol even when the snapshots are trivial. The role of time-glide symmetry in these transitions is also elucidated.

In Sec. \ref{s2}, we start by presenting our Hamiltonian model. We derive analytical results using the Floquet operator formalism in Sec. \ref{s3}. Section \ref{s4} presents symmetry arguments and relevant topological invariants. Then we follow with a discussion of the bulk properties using the replica picture based on solving the Floquet Hamiltonian in Sec. \ref{s5}. Section \ref{s6} is devoted to our numerical results for finite systems highlighting topological states. Also, topological phase diagrams are investigated in Sec. \ref{s7}. Finally, Sec. \ref{s8} collects our final remarks.
\begin{figure}[t!]
    \centering
    \includegraphics[width=1\linewidth]{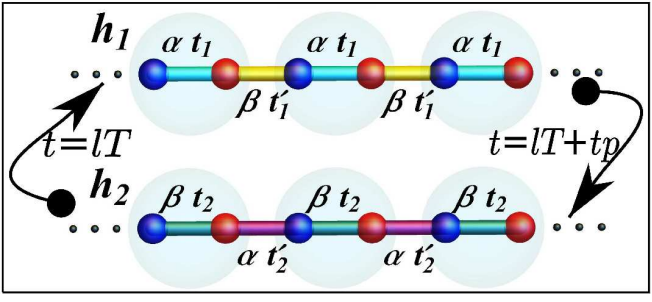}
    \caption{(Color online) Schematic illustration of the periodically quenched SSH lattice model over a complete driving period $T$ with two time durations $t_p$ and $T-t_p$. Each unit cell contains two sublattices A and B shown, respectively, by blue and red spheres.}
    \label{fig1}
\end{figure}

\section{Model}\label{s2}

We consider SSH model under piecewise periodic quenched dimerization as shown in Fig. \ref{fig1}. The Floquet evolution can be generated by a time-dependent Hamiltonian $H(t)$;  
\begin{eqnarray}\label{Eq1}
H(t)=   
    \begin{cases} 
      h_1, \quad\text{if} \quad \quad \quad\ lT\leq t<t_p+lT \\
      h_2, \quad \text{if} \quad\ t_p+lT\leq t<T+lT
    \end{cases},
\end{eqnarray}
as
\begin{eqnarray}\label{Eq2}
U(T)=e^{-i h_2(T-t_p)/\hbar}e^{-i h_1t_p/\hbar},
\end{eqnarray}
where the driving period $T$ is decomposed into two arbitrary time durations $t_p$ and $T-t_p$, and $l \in \mathbb{Z}$. The Hamiltonians in the two time durations are
\begin{eqnarray}\label{Eq3}
h_j &=& \alpha^{\delta_{j,1}}\beta^{\delta_{j,2}} \sum_{i=1}^N t_j A_i ^\dagger B_i\\ 
&+&\beta^{\delta_{j,1}}\alpha^{\delta_{j,2}} \sum_{i=1}^{N-1} t_j^{\prime}(A_{i+1} ^\dagger B_{i})+h.c,\nonumber
\end{eqnarray}
where $j=1,2$, the intra unitcell and inter unitcell hoppings, respectively, are $t_j=1+\Delta t_j$ and $t_j^{\prime}=1-\Delta t_j$ with $\Delta t_1=\Delta_0 \cos(\theta)$ and $\Delta t_2=\Delta_0 \cos(\theta+\phi)$ are the strength of dimerization for the two time durations with the amplitude $\Delta_0=0.8$ and phase $\theta$. $\phi$ is the phase shift of dimerization between the two time durations, and $\delta_{i,j}$ is the Kronecker delta function. Also, the parameters $\alpha$ and $\beta$ allow to control the relative magnitude of the hoppings on each piece of the driving protocol. Note that for the $\phi=0$, the dimerizations of the two time durations are similar to each other while for $\phi=\pi$, they have opposite dimerizations. 
In the presence of transnational symmetry, the Bloch wavevector $k$ is a good quantum number, and one can take Fourier transformation yielding the time-dependent Bloch Hamiltonian $H(k,t)$ as
\begin{eqnarray}\label{Eq4}
H(k,t)=
\begin{cases} 
       h_1(k), \quad\text{if}\quad \quad \quad\ lT\leq t<t_p+lT \\
      h_2(k), \quad \text{if}\quad\ t_p+lT\leq t<T+lT
    \end{cases},
\end{eqnarray}
where 
\begin{eqnarray}\label{Eq5}
h_{j}(k)&=&h_{jx}(k)\sigma_x+h_{jy}(k)\sigma_y,
\end{eqnarray}
with
\begin{eqnarray}\label{Eq6}
h_{1x}(k)&=&\alpha t_1 + \beta t_1^{\prime} \cos(k), \quad h_{1y}(k)= \beta t_1^{\prime}\sin(k), \nonumber \\ 
h_{2x}(k)&=& \beta t_2 + \alpha t_2^{\prime} \cos(k),  \quad h_{2y}(k)= \alpha t_2^{\prime} \sin(k).
\end{eqnarray}
Here, $\sigma_x$ and $\sigma_y$ are the Pauli matrices.

\section{Analytical results based on the Floquet operator formalism}\label{s3}
 
Since the stroboscopic dynamics of a Floquet system is governed by the time-evolution operator over a complete driving period, we introduce the time evolution operators of the two durations as $u_1=e^{-i h_1(k)t_p/\hbar}$ and $u_2=e^{-i h_2(k)(T-t_p)/\hbar}$. So the Floquet operator, i.e., $U(T,k)=\mathcal{T} \int_{t_0}^{t_0+T} e^{-iH(k,t)t/\hbar}dt$, can be written as
\begin{eqnarray}\label{Eq7}
U(T,k)\!&=&\!u_2u_1\nonumber\\
\!&=&\!e^{-i (h_{2x}\sigma_x + h_{2y}\sigma_y)(T-t_p)} e^{-i (h_{1x}\sigma_x+h_{1y}\sigma_y)t_p}.
\end{eqnarray}
Here, $\mathcal{T}$ enforces the time ordering, $t_0$ is an initial time, and we have taken $\hbar=1$. The $U(T,k)$ in Eq. (\ref{Eq7}) can be written as $ U(T,k) = e^{-i E(k) \hat{\boldsymbol{n}}\cdot \boldsymbol{\sigma}}$ in some special case where $E(k)$ is the quasi-energy, $\hat{\boldsymbol{n}}$ is a unit vector and $ \boldsymbol{\sigma} = (\sigma_x, \sigma_y, \sigma_z)$ is the Pauli vector. In the SSH model, the topological phase transition occurs at $k_s=0$ and $k_s=\pm\pi$ \cite{SSH,SSH1,GeSSH,GeSSH0,SSHZeeman,GeSSH1}. The Floquet operator at these points is
\begin{eqnarray}\label{Eq8}
U(T,k_s)&=&e^{-i E(k_s) \sigma_x}\nonumber\\
&=&\left(\begin{array}{cc}
 \cos(E(k_s))& -i\zeta \sin(E(k_s))\\ 
 -i\zeta \sin(E(k_s))&\cos(E(k_s))
 \end{array}\right),
\end{eqnarray}
with
\begin{eqnarray}\label{Eq9}
E(k_s)=E_2(k_s)(T-t_p)+\xi E_1(k_s)t_p,
\end{eqnarray}
and 
\begin{eqnarray}\label{Eq10}
E_j(k_s)=|\alpha^{\delta_{j,1}}\beta^{\delta_{j,2}}t_j+e^{ik_s}\beta^{\delta_{j,1}}\alpha^{\delta_{j,2}}t_j^{\prime}|,
\end{eqnarray}
where $\zeta=sgn(\beta t_2-\alpha t_2^{\prime})$ with $sgn$ being the Sign function. Also, $E_j(k_s)$ is the eigenvalue of $h_j(k)$ evaluated at the $k_s$ and
\begin{eqnarray}\label{Eq11}
\xi=\begin{cases}
       +1 \quad if \quad sgn(\alpha t_1-\beta t_1^{\prime})=sgn(\beta t_2-\alpha t_2^{\prime}) \\
       -1 \quad if \quad sgn(\alpha t_1-\beta t_1^{\prime})\ne sgn(\beta t_2-\alpha t_2^{\prime})
\end{cases}\!\!\!,
\end{eqnarray}
where $\xi=+1$ means that the two time durations have the same topology, but $\xi=-1$ means that the topology of each time duration differs from the other one. This is anticipated from the static SSH model where if the inter unitcell hopping is larger (smaller) than the intra unitcell hopping, the system hosts a  nontrivial (trivial) topological phase \cite{SSH,SSH1}.

Using the relation $ e^{i\epsilon\hat{\boldsymbol{n}}\cdot\boldsymbol{\sigma}} = \cos(\epsilon) I + i \sin(\epsilon) \hat{\boldsymbol{n}}\cdot\boldsymbol{\sigma}$, where $I$ is an identity matrix of size 2, in the first line of Eq. (\ref{Eq8}), one can see that $E(k_s)$ satisfies the equation $\cos(E(k_s))=\cos(\theta)$. Then if the gap closes at the center of Floquet-BZ ($\epsilon=0$) or at the edge of the Floquet-BZ ($\epsilon=\pm \pi/T$), i.e., $E(k_s) = 0$ or $E(k_s) = \pm \pi/T$, we will have $\cos(E(k_s))=1$ or $\cos(E(k_s))=-1$, respectively. So, the energy gap closure conditions can be obtained as
\begin{eqnarray}\label{Eq12}
E_1(k_s)&=&\xi \frac{\epsilon-(T-t_p)E_2(k_s)+2\pi n}{t_p},
\end{eqnarray}
where $n \in \mathbb{Z}$. Inserting Eq. (\ref{Eq10}) into the above equation, the topological phase transition can be determined as
\begin{eqnarray}
\Delta t_1&=&\xi(\frac{\epsilon-(T+(-1+\xi)t_p)(\alpha+\beta)+2\pi n}{t_p(\alpha-\beta)} \nonumber \\
&+&\frac{\Delta t_2(T-t_p)}{t_p})\quad  if \quad \alpha \ne \beta,  \label{Eq13} \\
\alpha&=&\beta=-\frac{\epsilon+2\pi n}{2(T+(\xi-1)t_p)},  \label{Eq14} 
\end{eqnarray} 
at $k_s=0$. In addition, for $\xi=-1$ there is another gap closing condition at $k_s=0$ with $\epsilon=0$:
\begin{eqnarray} \label{Eq15} 
T=2t_p \quad if \quad \beta=\alpha\frac{(\Delta t_2-1)T+4t_p}{(1+\Delta t_2)T}.
\end{eqnarray}
Note that if $\alpha=\beta$, the topological phase transition at $k_s=0$ is independent of the strengths of dimerization, i.e., $\Delta t_1$ or $\Delta t_2$. Also, a topological phase transition can also be occurred at $k_s=\pm\pi$:
\begin{eqnarray} \label{Eq16} 
\Delta t_1&=&\xi(\frac{\epsilon+(T-(1+\xi)t_p)(\alpha-\beta)+2\pi n}{t_p(\alpha+\beta)} \nonumber \\
&-&\frac{\Delta t_2(T-t_p)}{t_p}).
\end{eqnarray}
In addition, for $\xi=+1$, there is another gap closing condition at $k_s=\pm\pi$ with $\epsilon=0$:
\begin{eqnarray} \label{Eq17} 
T=2t_p \quad if \quad \beta=-\alpha\frac{(\Delta t_2-1)T+4t_p}{(1+\Delta t_2)T}.
\end{eqnarray} 

According to Eqs. (\ref{Eq13})-(\ref{Eq17}), there are two possibilities for gap closing taking place at either $k_s=0$ or $k_s=\pm\pi$ such that if the gap remains closed at a one of $k_s$'s, the topological phase transition would be occurred at the other one. For instance, if we set $\alpha=\beta=1$, the gap of system around $\epsilon=0$ for $k_s=0$ is always close while for $k_s=\pm\pi$ it would be closed when $\Delta t_1=0$. So, we can say that the topological phase transition can take place at $k_s=\pm\pi$ while the gap of system is always close at the other $k_s$ point, i.e., $k_s=0$. 

In the following, we will focus on discussing about three special cases: $i$) $\phi=0$, $ii$) $\phi=\pi$, and iii) $\beta=0$. The case $\phi=0$, i.e., $\Delta t_1=\Delta t_2$, resembles to a time-independent system effectively, if $\alpha=\beta$. Moreover, for case $\phi=0$ the topological phase transition reduces as
\begin{eqnarray}
\Delta t_1&=&\frac{\epsilon\!+\!(T+\xi(e^{ik_s}-\xi)t_p)(\alpha\!+\!e^{ik_s}\beta)\!+\!2\pi  n}{(T-\xi(e^{ik_s}+\xi)t_p)(\alpha-e^{ik_s}\beta)}\! \nonumber \\ 
&if& \!\quad (T-\xi(e^{ik_s}+\xi)t_p)(\alpha-e^{ik_s}\beta)\ne 0, \label{Eq18} \\
      \alpha&+&\xi \beta=\xi(e^{ik_s}+\xi)\frac{ \epsilon+2\pi n}{2T} \nonumber \\ 
      &if& \quad \begin{cases}
             t_p=T/2, \\
             \alpha=\frac{\epsilon+2\pi n}{2(T+\xi(e^{ik_s}-\xi)t_p)}\!\quad if \!\quad T \ne 2t_p.
      \end{cases}   \label{Eq19}
\end{eqnarray}

For the  case $\phi=\pi$, i.e., $\Delta t_1=-\Delta t_2$, the condition of topological phase transitions can be obtained as
\begin{eqnarray}
\Delta t_1&=&\frac{\epsilon\!-\!(T+\xi(e^{ik_s}-\xi)t_p)(\alpha\!+\!e^{ik_s}\beta)\!+\!2\pi  n}{(T+\xi(e^{ik_s}-\xi)t_p)(\alpha-e^{ik_s}\beta)}\! \nonumber \\ 
&if& \!\quad (T+\xi(e^{ik_s}-\xi)t_p)(\alpha-e^{ik_s}\beta)\ne 0,\label{Eq20} \\
\alpha&=&e^{ik_s} \beta=\frac{\epsilon+2\pi n}{2(T+\xi(e^{ik_s}-\xi)t_p)}\label{Eq21} \\ 
t_p&=&\delta_{\xi,e^{i(k_s-\pi)}}T/2 \quad if \quad \epsilon=0.\label{Eq22}
\end{eqnarray}
According to these relations, the gap around $\epsilon=0$ can be closed at both $k_s=0$ and $\pi$ at the same time. As shown in Eq. (\ref{Eq22}), for $\phi=\pi$ whenever $t_p=T/2$ the gap of system around zero quasi-energy is always close. We will show below that this point is a symmetric time point for time-glide symmetry.

For the other special case $\beta=0$, according to Eq. (\ref{Eq6}), the system reduced to a system that only has an intra (inter) unitcell hopping for the first (second) time duration resulting in the case where the two time durations have opposite topological phase, i.e., $\xi=-1$. Furthermore, the topological phase transitions  (\ref{Eq18}) and (\ref{Eq19}) for $\phi=0$ reduce as \begin{eqnarray}
\Delta t_1&=&\frac{\epsilon\!+\!(T-(e^{ik_s}+1)t_p)\alpha\!+\!2\pi  n}{(T+(e^{ik_s}-1)t_p)\alpha}\! \nonumber \\ 
&if& \!\quad t_p \ne T/2, \label{Eq23} \\
\alpha&=&\delta_{-1,e^{i k_s}}\frac{\epsilon+2\pi n}{T} \quad if \quad t_p=T/2. \label{Eq24}
\end{eqnarray}
The above equation shows that for $t_p=T/2$, the gap around zero ($\pm \pi/T$) quasi-energy is always close at $k_s=\pm\pi$ if $\alpha=n \omega$ ($\alpha=n\omega/2$). Although, the topological phase transition can be occurred at the other supersymmetry point. Also, for $\beta=0$ and $\phi=\pi$ the topological phase transitions (\ref{Eq20})-(\ref{Eq22}) reduce as
\begin{eqnarray}
\Delta t_1&=&\frac{\epsilon\!+\!(T-(e^{ik_s}+1)t_p)\alpha\!+\!2\pi  n}{(T-(e^{ik_s}+1)t_p)\alpha}\! , \label{Eq25} \\
t_p&=&\delta_{-1,e^{i(k_s-\pi)}}T/2 \quad if \quad \epsilon=0. \label{Eq26}
\end{eqnarray}
Also, the above equation shows that for the case $\phi=\pi$ the gap of system is always close at zero quasi-energy and at $k_s=0$ if $t_p=T/2$. 

\section{Symmetry arguments and topological invariants}\label{s4}

A Floquet system described by the Floquet operator $U(k)$ has chiral symmetry if there exists a unitary transformation $\Gamma$, such that $\Gamma^2 = 1$ and $\Gamma U_i \Gamma = U_i^{-1}$ with $i= 1, 2$ where $U_{1,2}$ are Floquet operators in two different time durations \cite{Floq1D0,Floq1D1}. As such, the chiral symmetry for the Floquet operator $U(k)$ leads to the existence of a pair of symmetric time durations, which can be related to each other by evolving the time. So, the Floquet operators in the two different time durations can then be obtained as $U_1 = u_1u_2u_1$ and $U_2 = u_2u_1u_2$ by shifting the starting time with $\tau_1=t_p/2$ and $\tau_2=(T-t_p)/2$. Also, the product of $u_1$ and $u_2$ generates the Floquet operator $U(k)$ as $U(k) = u_2u_1$. The Floquet operators in the symmetric time durations can then be obtained as
\begin{eqnarray}
U_1&=&u_1(\tau_1)u_2(2\tau_2)u_1(\tau_1) \nonumber \\ &=&\left (\begin{array}{cc}
 n_{1z}&n_{1x}-in_{1y}\\ 
 n_{1x}+in_{1y}&n_{1z}
 \end{array}\right),\label{Eq27}\\
U_2&=&u_2(\tau_2)u_1(2\tau_1)u_2(\tau_2) \nonumber \\
&=&\left(\begin{array}{cc}
 n_{2z}&n_{2x}-in_{2y}\\ 
 n_{2x}+in_{2y}&n_{2z}
 \end{array}\right).\label{Eq28}
\end{eqnarray}
In our system the chiral operator is $\Gamma=\sigma_z$. Therefore, following Refs. \cite{Floq1D0,Floq1D1}, to define topological invariants of chiral-symmetric Floquet systems, one can introduce a pair of winding numbers 
\begin{eqnarray}\label{Eq29}
\nu_\eta=\frac{1}{2 \pi}\int_{-\pi}^{\pi} \frac{n_{\eta x}(k) \partial_k n_{\eta y}(k) - n_{\eta y}(k) \partial_k n_{\eta x}(k)}{n_{\eta x}^2(k) + n_{\eta y}^2(k)},
\end{eqnarray}
where $\eta=1,2$ stand for the winding numbers of Floquet operators $U_1(k)$ and $U_2(k)$.
By adding and subtracting the winding numbers $(\nu_1, \nu_2)$, another pair of winding numbers can be obtained as \cite{Floq1D0,Floq1D1}
\begin{eqnarray}\label{Eq30}
\mathrm{W}_0=\frac{\nu_1 + \nu_2}{2}, \quad \mathrm{W}_{\pi}=\frac{\nu_1 - \nu_2}{2},
\end{eqnarray}
which could fully determine topological properties of bulk Floquet states of $U(k)$. Explicitly, these winding numbers take nontrivial values for systems possessing chiral symmetry. Subsequently, there could be two band gaps at the $\epsilon=0$ and $\epsilon=\pm \pi/T$ quasi-energies. Under open boundary conditions, topologically protected edge states could appear in the gaps. The integer-quantized topological invariants $\mathrm{W}_0$ and $\mathrm{W}_\pi$ defined in Eq. (\ref{Eq30}) enumerate exactly the number of degenerate edge states at the center and at the edge of Floquet-BZ, respectively.

Now, we uncover a symmetry that may exist in the time domain. In the case $\phi=\pi$, according to Eqs. (\ref{Eq22})-(\ref{Eq26}) the gap of system is always close at $k_s=\pm\pi$ if $t_p=T/2$. This time point is a time symmetric point that can establish time-glide symmetry \cite{time-glide}. The time-glide symmetry is similar to reflection-glide symmetry \cite{reflection-glide1,reflection-glide2,reflection-glide3} where translation in space is replaced by translation in time. If $\phi=\pi$, the $H(k,t)$ has time-glide symmetry defined as 
\begin{eqnarray}\label{Eq31}
G H(k,t) G^{-1} = H(R(k),t+T/2),
\end{eqnarray}
and also 
\begin{eqnarray}\label{Eq32}
G h_1(k) G^{-1} = h_2(k),
\end{eqnarray}
where $R$ is reflection and the time-glide symmetry operator $G$ is
\begin{eqnarray}\label{Eq33}
G =\left (\begin{array}{cc}
 0&e^{ik/2}\\ 
 e^{-ik/2}&0
 \end{array}\right).
\end{eqnarray}
The time-glide symmetry enforces the gap to be closed at $k_s=\pm\pi$ and at zero energy. Changing the $t_p$ so that $t_p\ne T/2$ breaks such symmetry and opens a gap. Note, another gap closing can takes place at $k_s=0$ in the case $\phi =\pi$ being not related to time-glide symmetry. On the other hand, the $T$ is responsible for the gap closure at $k_s=0$. To more clarify the role of $t_p$ and $T$ on the band gap of system, and also, to explore the root of gap closing, we investigate the bulk band structure of system by employing the Floquet Hamiltonian below.


\section{Bulk band structures based on the Floquet Hamiltonian formalism}\label{s5}


A periodically driven quantum system can be described by a time-periodic Hamiltonian as
\begin{eqnarray}\label{Eq34}
H(t)=H(t+T).
\end{eqnarray}
Such system can have generalized stationary states $\vert \psi_n(t) \rangle$ called Floquet states \cite{FloState}. To obtain these states, one has to solve the time-dependent Schr\"{o}dinger equation
\begin{eqnarray}\label{Eq35}
H(t) \vert \psi_n(t) \rangle = i\partial_t \vert \psi_n(t) \rangle,
\end{eqnarray}
yielding,
\begin{eqnarray}\label{Eq36}
\vert \psi_n(t) \rangle =  \vert u_n(t) \rangle e^{-iE_n t},
\end{eqnarray}
where $E_n$ is the quasi-energy and the Floquet mode $\vert u_n(t) \rangle$ is time-periodic;
\begin{eqnarray}\label{Eq37}
\vert u_n(t) \rangle =  \vert u_n(t+T) \rangle.
\end{eqnarray}
The quasi-energies can be restricted to a first Floquet-BZ, in analogy with Bloch's theorem.
Plugging the Floquet states (\ref{Eq36}) into the time-dependent
Schr\"{o}dinger equation (\ref{Eq35}), one gets
\begin{eqnarray}\label{Eq38}
H_F\vert u_{n}(t) \rangle = E_{n}\vert u_{n}(t) \rangle.
\end{eqnarray}
where $H_F=H-i\frac{\partial }{\partial t}$ is the Floquet Hamiltonian. The matrix elements of the Floquet Hamiltonian $H_F$ can be written as,
\begin{equation}\label{Eq39}
H_F^{n,m}=\frac{1}{T}\int _{0}^{T} H(t) e^{i(n-m)\omega t}dt - n \omega \delta_{n m},
\end{equation}
where $\omega=2\pi/T$ is the frequency.

Substituting Eq. (\ref{Eq4}) into Eq. (\ref{Eq39}), one can obtain the Floquet Hamiltonian of the system as  
\begin{eqnarray}\label{Eq40}
H_F^{n,m}&=&\frac{1}{T}(\int _{0}^{t_p}\!\!\!\! h_1 e^{i(n-m)\omega t}dt+\!\!\int _{t_p}^{T} \!\!\!\!h_2 e^{i(n-m)\omega t}dt) - n \omega \delta_{n m},\nonumber\\
\end{eqnarray}
simplifying to 
\begin{eqnarray}\label{Eq41}
H_F^{n,m}=\begin{cases}
        \frac{-i}{2 \pi \omega (n-m)}[(e^{i (n-m)t_p \omega}-1)h_1(k) \\
        + (e^{2\pi i (n-m)}-e^{i(n-m)t_p \omega}) h_2(k)] \quad if \quad n \ne m, \\
        \\
        \frac{h_1(k)t_p+h_2(k)(T-t_p)}{T}-n\omega \quad if \quad n=m.
        \end{cases}
\end{eqnarray}
To proceed further, we need to truncate the Floquet space. The matrix structure of Floquet Hamiltonian can be written as 
\begin{eqnarray}\label{Eq42}
H_F&=&\left(\begin{array}{ccc}
 H^{-1,-1}+\omega & H^{0,-1} &  H^{1,-1}\\ 
 H^{-1,0} & H^{0,0} &  H^{1,0}\\ 
  H^{-1,1} & H^{0,1} &  H^{1,1}-\omega
 \end{array}\right),
\end{eqnarray}
up to the first order. The band structure comprising of quasi-energy allows us to determine the dynamics of periodically driven system. Note that the non-vanishing elements $H^{n,m}$ with $n\neq m$ can couple the $n$ and $m$ replicas together. Generally, for Floquet operator formalism, because of the brunch cut of logarithm, the first Floquet-BZ is $-\pi/T<E<\pi/T$, while for Floquet Hamiltonian one the first Floquet-BZ is $-\omega/2<E<\omega/2$. 

The Floquet Hamiltonian (\ref{Eq42}) has inversion symmetry, that is $\Pi H_F(k) \Pi=H_F(-k)$ where the inversion operator is $\Pi=\it{I}_n \otimes \sigma_x$ with $\it{I}_n$ being an identity matrix with dimensions given by the number of considered Floquet replicas $n$. This symmetry motivates us to use the multi-band Zak phase \cite{Zak} as a topological invariant to characterize the topological edge states with the $\epsilon=0$ and $\epsilon=\pm \omega/2$ quasi-energies. Note that, in some cases, the $n=0$ replica is responsible for inducing the edge state with the $\epsilon=0$ quasi-energy and it therefore suffices to use the block of Hamiltonian (\ref{Eq41}) associated with $n=0$ replica. For the study of the edge states with the $\epsilon=\pm \omega/2$ quasi-energy, one should use the whole Hamiltonian (\ref{Eq42}) invoking the $n=0$ and $n=1$ for $\epsilon= \omega/2$ edge states and $n=0$ and $n=-1$ for $\epsilon=-\omega/2$. Note also that a coupling between $n=0$ and $n=\pm1$ will produce the relevant gap opening around $\epsilon=\pm \omega/2$.

An advantage of this solution scheme is that it allows to separate the contributions from different replicas. In our context it allows us to distinguish the replicas playing a role on the formation of edge states and/or the continuum states that may coexist with them. For example, if we know that a bandgap is formed either on one replica or because of the hybridization between two replicas, it is expected that the midgap edge states (if they exist) also have a weight on the parent replicas, then by analyzing the weights on the remaining ungapped bands (probably at a different $k$ value), one can infer whether the edge states will hybridize with the continuum states or not. 

The weights of a given state $|\psi \rangle$ on replica $n$ is defined by:
\begin{eqnarray}\label{e43W}
\mathcal{W}_n=\sum_j |\langle j,n|\psi \rangle |^2,
\end{eqnarray}
where $|j,n \rangle$ is the basis of states localized at site $j$ and Floquet replica $n$ and the sum runs over all values of the site index $j$. 

Using Hamiltonian (\ref{Eq42}), we will determine the system bulk state behavior under periodic boundary conditions and examine what will happen for different parameters. In what follows, we consider the system in two scenarios. In the first scenario, we investigate the system for partial dimerization case, i.e., $\beta \ne 0$, resulting in topological phase transition in the presence of semimetal phase at the center and at the edge of Floquet-BZ. Also, we show the effect of $t_p$ and $T$ on the band structure and examine which gap is affected by $t_p$ and $T$. In the second scenario, each half of Hamiltonian has full dimerization, i.e., $\beta=0$, where the system can be considered as a time version of the SSH model for $\phi=0$.

\begin{figure}[t!]
    \centering
    \includegraphics[width=1\linewidth]{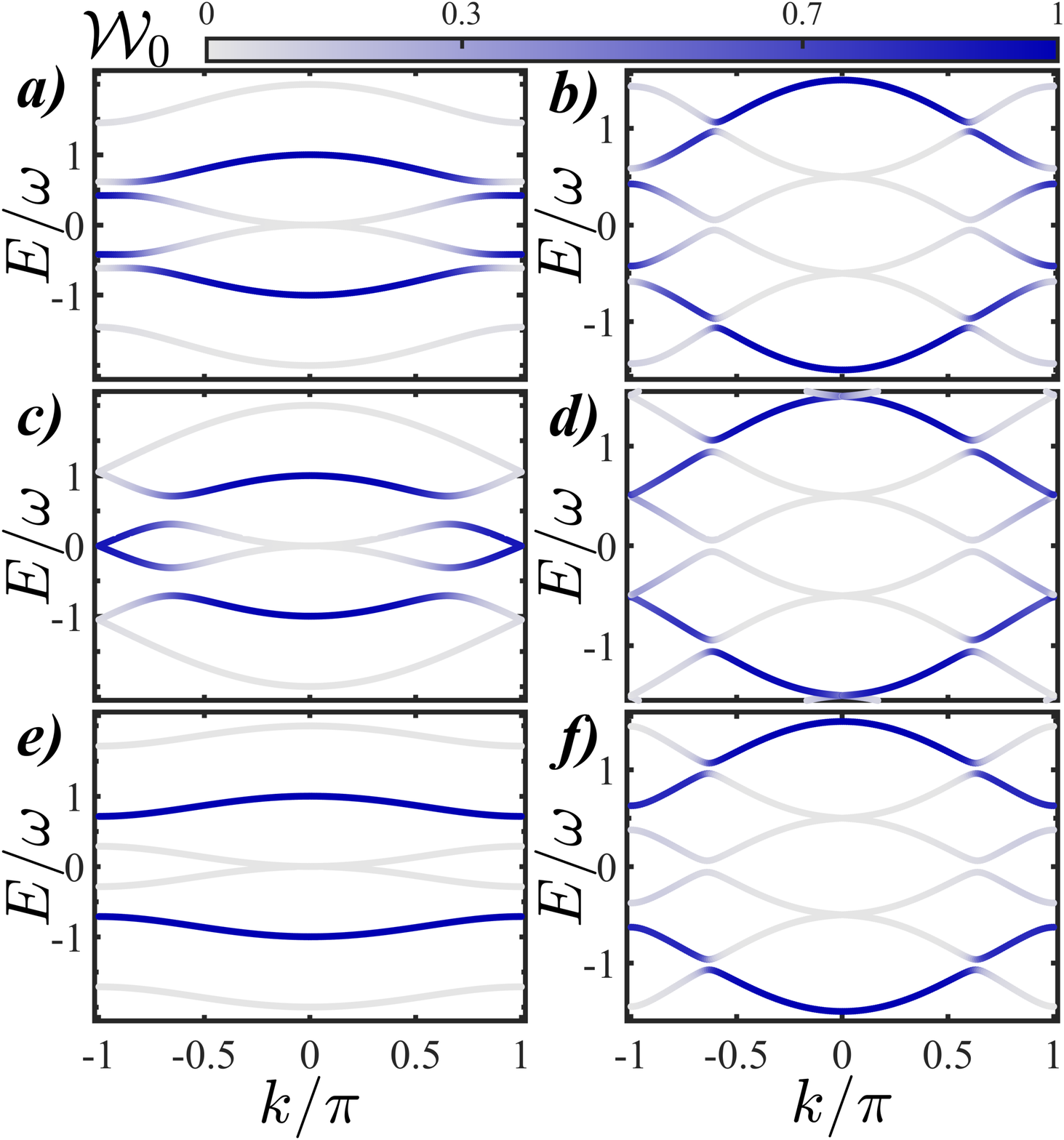}
    \caption{(Color online) Quasi-energy spectra versus $k$ with $\alpha=\beta=1$ and $\phi=0.3\pi$ and $t_p=0.5T$ under periodic boundary conditions. Top row: Before the topological phase transition. Middle row: At the topological phase transition. Bottom row: After the topological phase transition. Left column: $\omega=2$. Right column: $\omega=4/3$. Here, the color scale encodes the weight $\mathcal{W}_0$ on the $n=0$ replica. }
    \label{fig2}
\end{figure}
We start investigating the first scenario. The spectra as a function of $k$ for $\phi=0.3$ and $\alpha=\beta=1$ are shown in Fig. \ref{fig2} before (top row), at (middle row), and after (bottom row) the topological phase transition point to show the closing and reopening of the gap at the center (left column) and at the edge (right column) of Floquet-BZ with $k_s=\pm\pi$. One can see, for $\omega=2$ (the left column of Fig. \ref{fig2}), according to Eq. (\ref{Eq16}), a topological phase transition can be occurred by closing and reopening the zero energy gap of the Floquet replica $n=0$, indicated by blue color, at $k_s=\pm\pi$. Subsequently, the degeneracy points at the topological phase transition point with $k_s=\pm\pi$ (see Fig. \ref{fig2}(c)) are lifted before (see Fig. \ref{fig2}(a)) and after (see Fig. \ref{fig2}(e)) the topological phase transition. On the other hand, based on Eq. (\ref{Eq14}) the degeneracy point at $k_s=0$ with zero quasi-energy remains preserved. So there is no gap between the $n=\pm 1$ Floquet replicas, indicated by gray color, providing semimetal phase at zero energy. As a result, a topological phase associated with zero-energy edge states in the presence of semimetal states can be established (see also Fig. \ref{fig5}(a)). For $\omega=4/3$, as shown in the right column of Fig. \ref{fig2}, at the edge of Floquet-BZ ($\epsilon=\pm\omega/2$) and $k_s=\pm\pi$, according to Eqs. (\ref{Eq14}) and (\ref{Eq16}), the gap between replicas $n=\pm1$ is always close, while the gap of $n=0$ replica closes and then reopens. Subsequently, a topological phase transition is occurred at $k_s=\pm \pi$ and at the edge of Floquet-BZ. Moreover, the presence of semimetallic phase at such energies has been guaranteed by remaining closed the gap of $n=\pm1$ replicas at $k_s=0$. Consequently, one may anticipate a topological phase associated with $\pi$-mode within the semimetal bulk states (see also Fig. \ref{fig5}(b)).

\begin{figure}[t!]
    \centering
    \includegraphics[width=1\linewidth]{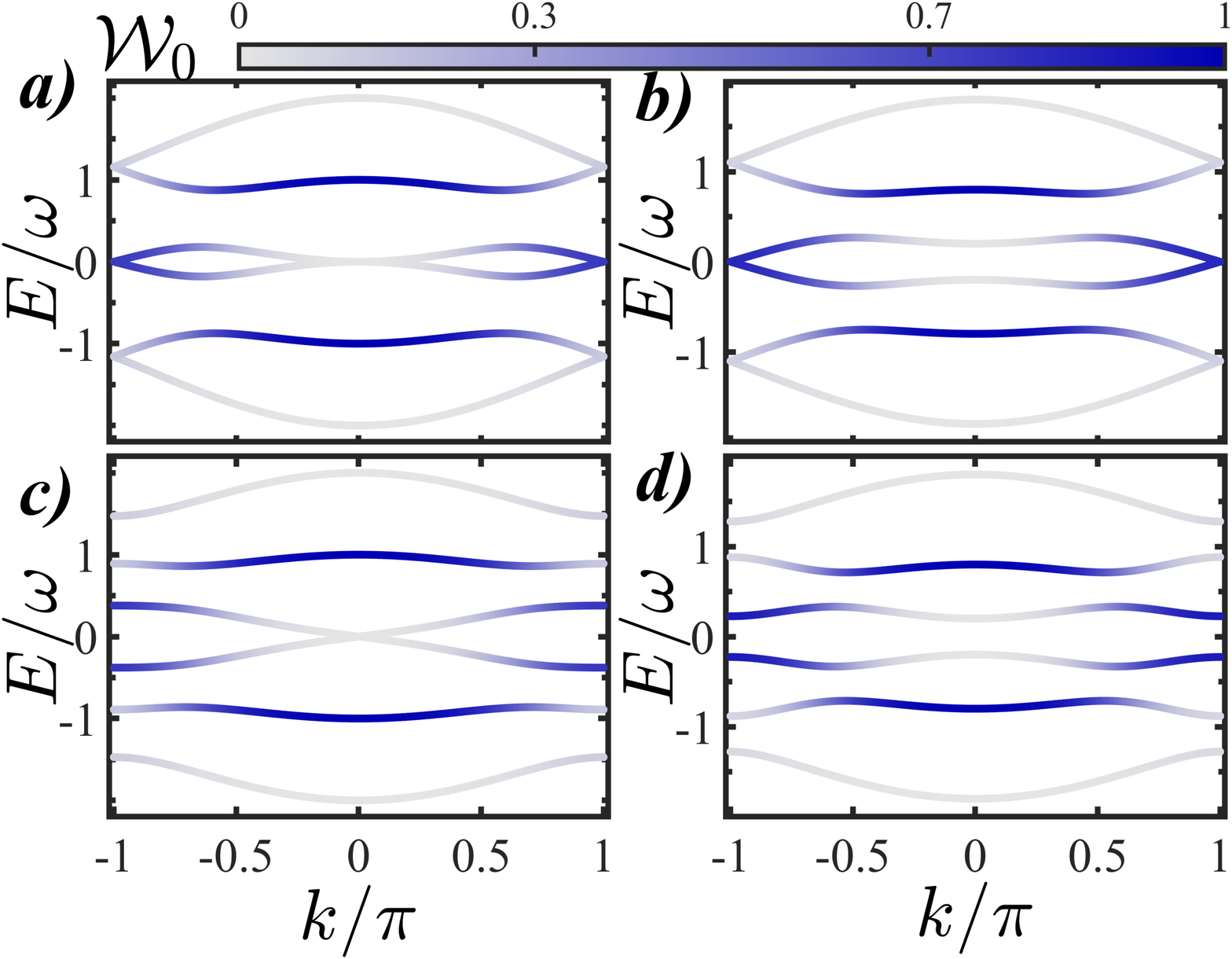}
    \caption{(Color online) Quasi-energy spectra versus $k$ with $\alpha=\beta=1$ and $\phi=\pi$ under periodic boundary conditions. Left column: $\omega=2$. Right column: $\omega=2.5$. Top row: $t_p=0.5T$. Bottom row: $t_p=0.3T$. Here, the color scale encodes the weight $\mathcal{W}_0$ on the $n=0$ replica.}
    \label{fig3}
\end{figure}
The dependence of quasi-energy spectra on $k$ is shown in Fig. \ref{fig3} for different values of $\omega$ (or equivalently $T$) and $t_p$ with $\alpha=\beta=1$ and $\phi=\pi$. In the top (bottom) row of Fig. \ref{fig3}, we set $t_p=0.5T$ ($t_p=0.3T$) and change $\omega=2$ to $\omega=2.5$ from the left to the right column. According to Eqs. (\ref{Eq20})-(\ref{Eq22}), one can see that the $\omega$ can control the gap between the $n=\pm 1$ replicas, indicated by gray color, at the state $k_s=0$ and at zero energy. The center of each replica is located at $\pm n\omega$, so for $\omega=2$ (left column), there is no gap between the two Floquet replicas $n=\pm1$ at topological phase transition point, but for $\omega=2.5$ (right column) the $n=\pm 1$ Floquet replicas are separated from each other and are gapped. In contrast, as shown in the top (bottom) row of Fig. \ref{fig3}, the change of $t_p$ from $t_p=0.5T$ to $t_p=0.3T$ violating the Eq. (\ref{Eq22}), causes that the $n=0$ Floquet replica becomes gapped at zero quasi-energy and at $k_s=\pm\pi$. In the case $\omega=2$, after closing and reopening the $n=0$ replica, corresponding to topological phase transition, if the system resides in a non-trivial regime, it would exhibit a hybridization between zero energy topological edge states and bulk states under open boundary condition (see also Fig. \ref{fig6}(c)). Because, there is a coupling between the $n=0$ and $n=\pm1$ replicas around the zero-quasi energies. Consequently, as depicted in Fig. \ref{fig3} (d), in which both $t_p$ and $\omega$ are changed simultaneously, all the bands are gapped and the system would show localized topological edge states under open boundary condition (see also Fig. \ref{fig6}(d)). Notice that for $\omega>2$, which means the $n=\pm1$ replicas are far away from each other, the gap is open without occurring any topological phase transition. While for $\omega<2$, a coupling between $n=\pm1$ replicas can be created and the system would host nontrivial topological phase. Furthermore, if we set $\omega<2$ and $t_p \ne 0.5T$ simultaneously, it is possible to establish non-trivial topological phase in the gaps of $n=0$ and $n=\pm1$ replicas resulting in the existence two pairs of edge states at the center and/or at the edge of Floquet-BZ.

\begin{figure}[t!]
    \centering
    \includegraphics[width=1\linewidth]{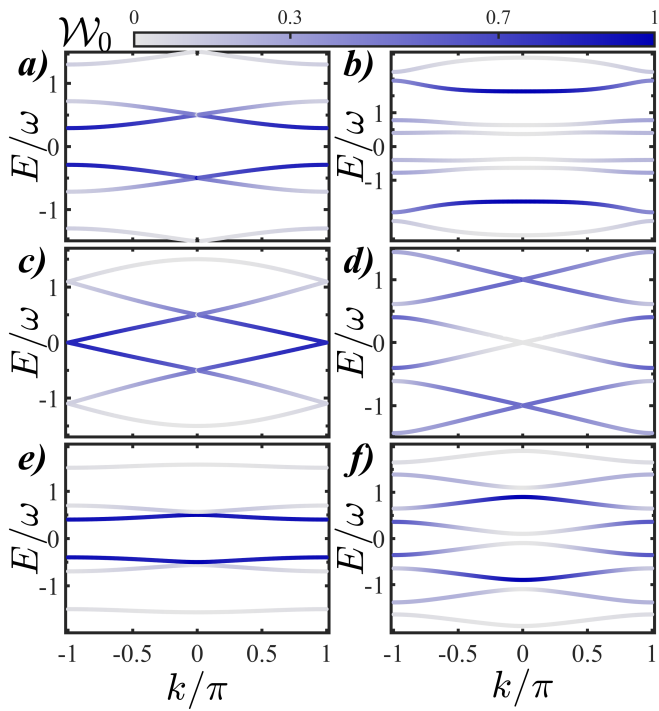}
    \caption{(Color online)  Quasi-energy spectra versus $k$ with $\alpha=1$ and $\beta=0$  under periodic boundary conditions. Top row: Before the topological phase transition. Middle row: At the topological phase point. Bottom row: After the topological phase transition. Left column: $\phi=0$, $t_p=0.5T$, and $\omega=2$. Right column: $\phi=\pi$, $t_p=0.3T$, and $\omega=1$. Here, the color scale encodes the weight $\mathcal{W}_0$ on the $n=0$ replica.}
    \label{fig4}
\end{figure}
The second scenario, where the Hamiltonian in each time duration has a full dimerization pattern, i.e., $\beta=0$, can reveal more interesting results. In particular, if $\phi=0$, the whole Hamiltonian includes partial dimerization in the time dimension illustrating a time version of SSH model. The quasi-energies in terms of $k$ are plotted in Fig. \ref{fig4} providing the gap closings and reopenings with $\phi=0$, $t_p=0.5T$, and $\omega=2$ for the left column and $\phi=\pi$, $t_p=0.3T$, and $\omega=1$ for the right column. As shown in Figs. \ref{fig4}(a), \ref{fig4}(c), and \ref{fig4}(e), at $k_s=\pm\pi$, according to Eqs. (\ref{Eq23})-(\ref{Eq24}), the gap of $n=0$ replica closes and then reopens representing a topological phase transition at zero energy which is the same for SSH model (see also Fig. \ref{fig5}(c)). However, the degenerate points at $k_s=0$ located on the edge of Floquet-BZ remain intact during the topological phase transition. 

In the full dimerization case if the patterns of two time durations are opposite, i.e., $\phi=\pi$, based on the first scenario and time-glide symmetry argument, one may expect that the gap of system at zero energy is always closed (see Fig. \ref{fig5}(d)). But, setting the $t_p$ away from the middle point breaks the time-glide symmetry. Therefore, more interestingly, as shown in Figs. \ref{fig4}(b), \ref{fig4}(d), and \ref{fig4}(f), according to Eq. (\ref{Eq25}), the gap of $n=\pm 1$ replicas closes and then reopens at zero energy and at $k_s=0$ resulting in the topological phase transition. As a result, when the intra-unitcell hopping in the first time duration is equal to the inter-unitcell hopping in the second time duration, i.e., $\beta=0$ and $\phi=\pi$, a nontrivial topological phase can be emerged in the system by changing the partition time $t_p$ (see also Fig. \ref{fig7}). To specify the region with topological edge states, we explore the quasi-energy and relevant topological invariant of Floquet operator in the following.

\section{Numerical results}\label{s6}
Using the Floquet operator (\ref{Eq2}), we will obtain the exact quasi-energy structure of the system and utilizing Eq. (\ref{Eq30}), the topological invariants can be evaluated numerically. We will also evaluate the normalized inverse participating ratio ($I_E$) \cite{IPR} for each eigenstate $\psi_n$ with eigenenergy $E_n$ to show the localization of the normalized eigenstates $\sum_n|\psi_n|^2=1$ given by $I_E=\sum_n|\psi_n|^4$.
Since localized states are unaffected by the boundaries, they are characterized by an $I_E$ independent of the system size. For localized states $I_E \rightarrow 1$ while for extended ones $I_E \rightarrow 0$. 

\begin{figure}[t!]
    \centering
    \includegraphics[width=1\linewidth]{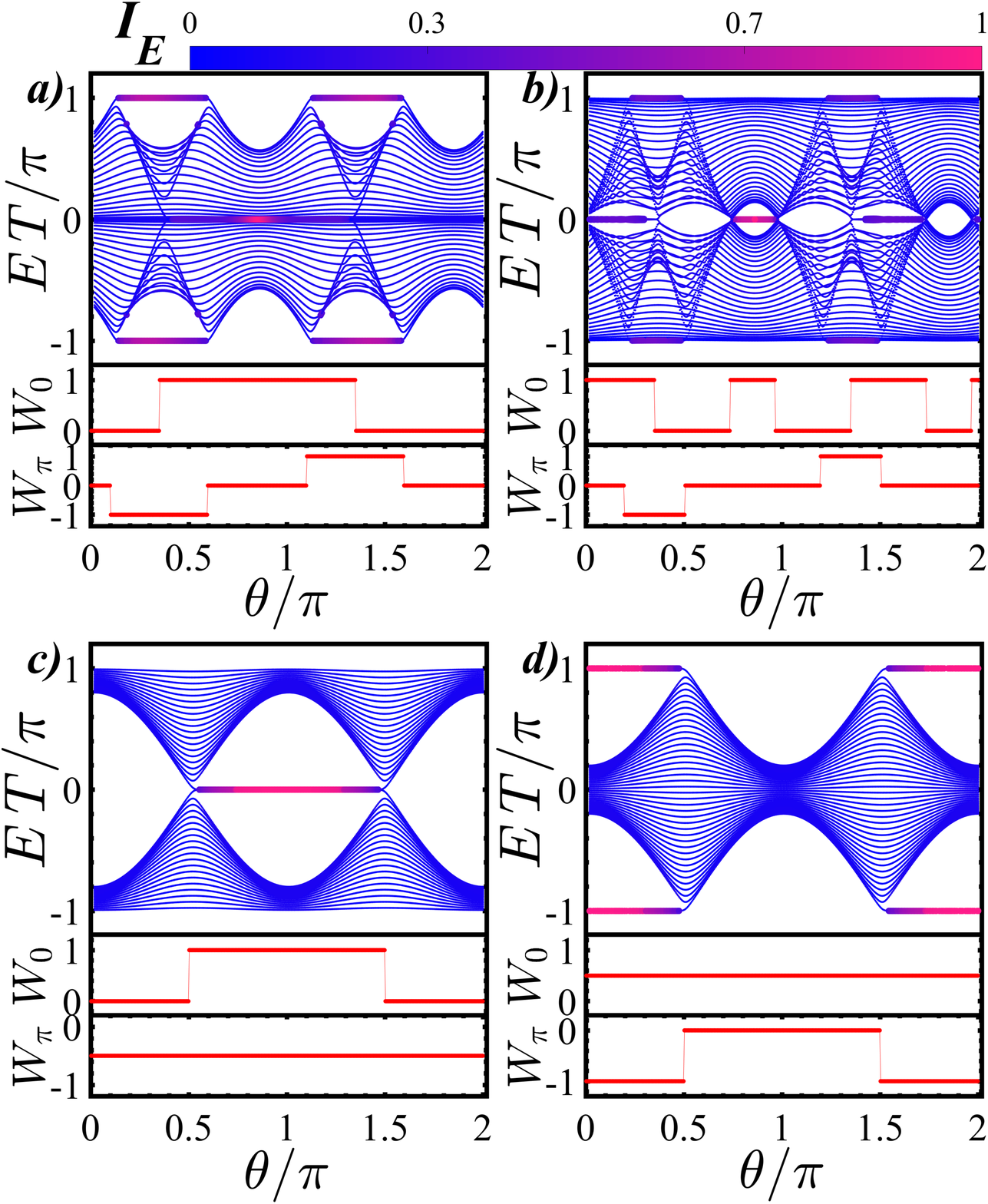}
    \caption{(Color online) Quasi-energy spectra and the relevant topological invariants as a function of $\theta/\pi$ for $t_p=0.5T$. In the top panels $\alpha=\beta=1$ and $\phi=0.3\pi$ with (a) $T=\pi$ and (b) $T=1.5\pi$, while the bottom panels have $\alpha=1$, $\beta=0$, and $T=\pi$ with (c) $\phi=0$ and (d) $\phi=\pi$.}
    \label{fig5}
\end{figure}

\begin{figure}[t!]
    \centering
    \includegraphics[width=1\linewidth]{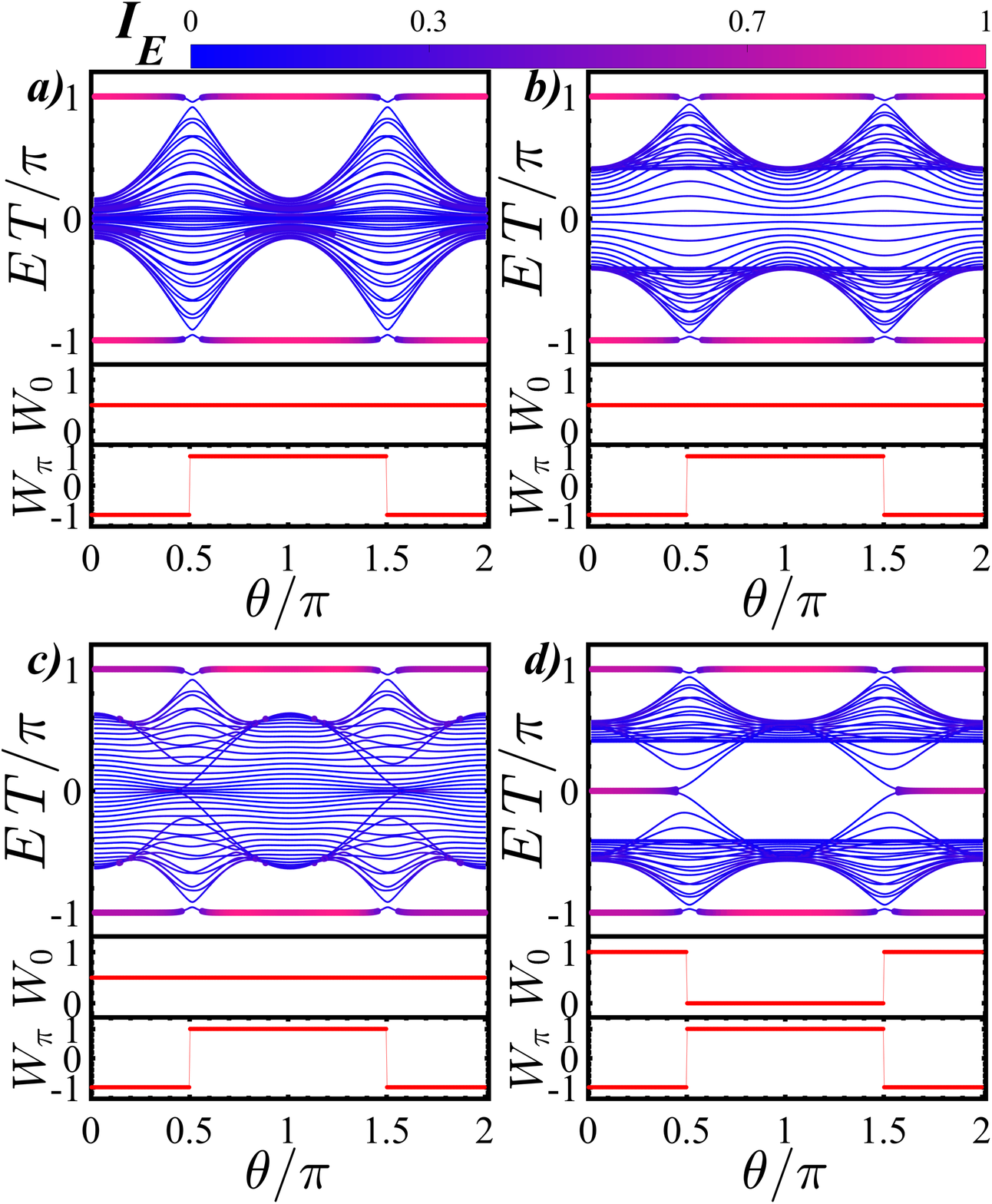}
    \caption{(Color online) Quasi-energy spectra and relevant topological invariant as a function of $\theta/\pi$ for $\alpha=\beta=1$, $\phi=\pi$, to show the influence of shifting the partition time and periodicity away from $t_p=T/2$ and $T=\pi$. For the top (bottom) panels $t_p=T/2$ ($t_p=0.3T$), while for the left (right) panels $T=\pi$ ($T=0.8\pi$).} 
    \label{fig6}
\end{figure}

\begin{figure}[t!]
    \centering
    \includegraphics[width=1\linewidth]{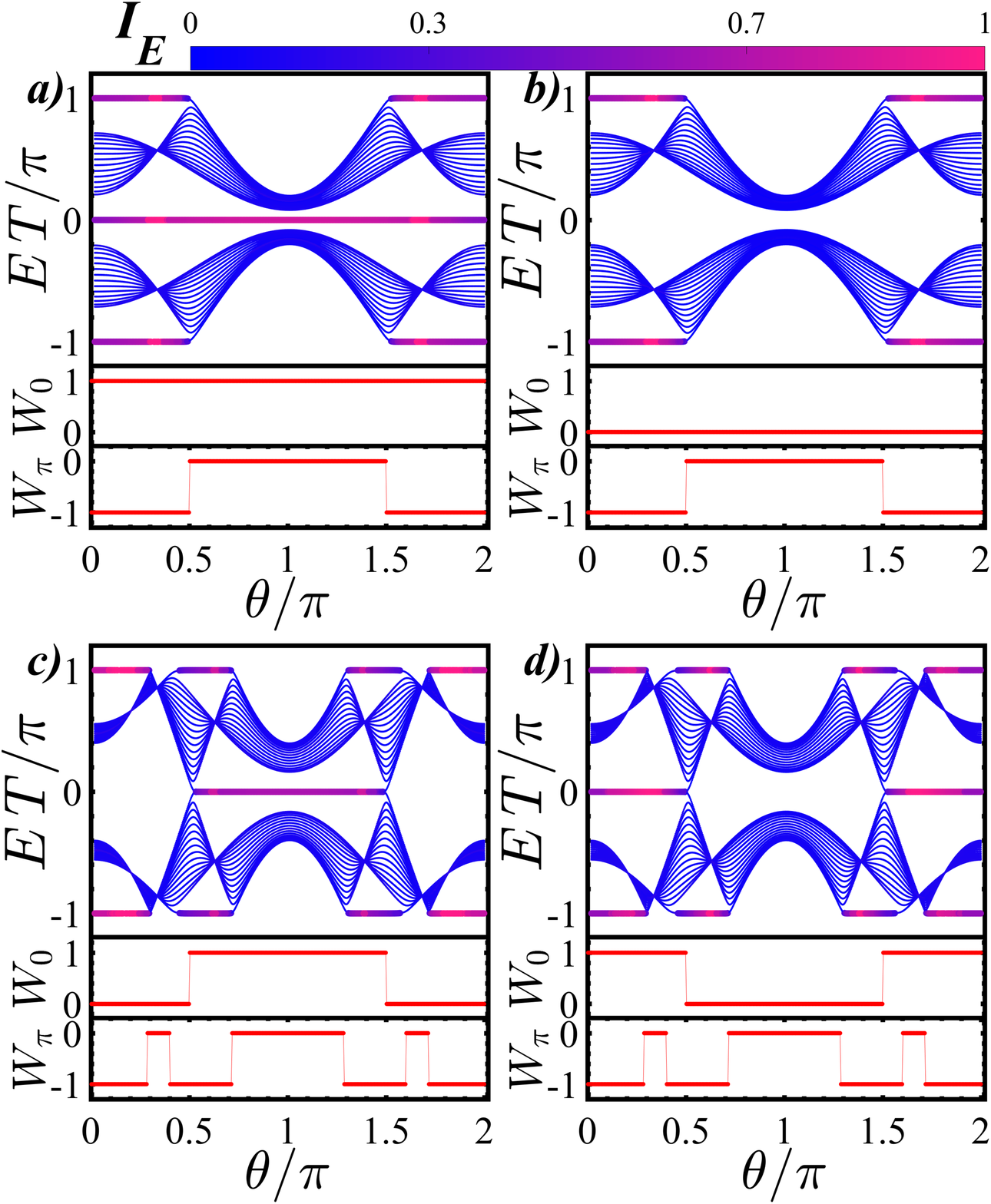}
    \caption{(Color online) Quasi-energy spectra and relevant topological invariant as a function of $\theta/\pi$ for $\alpha=1$, $\beta=0$, and $\phi=\pi$. The top (bottom) panels correspond to $T=\pi$ ($T=2\pi$), while the left (right) panels have $t_p=0.3T$ ($t_p=0.7T$).}
    \label{fig7}
\end{figure}

\subsection{confirming the analytical results} 

To confirm the above obtained analytical results and show the effect of existence or absence of time-glide symmetry, we have calculated the quasi-energy and the winding numbers as a function of $\theta/\pi$, shown in Figs. (\ref{fig5})-(\ref{fig7}). 

In Fig. (\ref{fig5}), the two top (bottom) panels have plotted for $\alpha=\beta=1$ and $\phi=0.3\pi$ ($\alpha=1,\beta=0$) for different $T$ ($\phi$). According to Eq. (\ref{Eq16}), the topological phase transition can be occurred for given parameters $T=\pi$ and $t_p=T/2$ at $\Delta t_1=-\Delta t_2$ and $k_s=\pi$. Also, according to Eq. (\ref{Eq14}), the condition that the gap remains closed is always fulfilled at $k_s=0$ around zero energy $\epsilon=0$. For such situation, as can be seen from Fig. \ref{fig5}(a), the zero-energy topological edge states coexist with the gapless bulk states \cite{TopoMetal1,TopoMetal3} providing the topological phase with the presence semimetal phase in the background. By choosing $T=1.5\pi$ ($\omega=4/3$), as already discussed above, the gaps around $\epsilon=\pm\pi/T$ quasi-energies are always closed at $k_s=0$ and, interestingly, as shown in Fig. \ref{fig5}(b), the $\pi$-mode edge states being induced in the system due to the closure of gap at $k_s=\pi$, coexist with the bulk states. This results in the existence of topological $\pi$-edge states in the semimetal spectrum. In addition, the zero quasi-energy gap is open and hosts zero-energy edge states. In Fig. \ref{fig5}(c), $\alpha=1$, $\beta=0$, and $\phi=0$, so the intra unitcell and inter unitcell hoppings are nonzero for the first and second time durations, respectively, imposing full dimerization pattern in each time duration. As already mentioned, this model resembles a time version of the original static SSH model~\cite{SSH}. As shown in the figure, the band structure of the model is the same as the static SSH one, having zero-energy topological edge states, without $\pi$-mode edge states. Because, based on Eq. (\ref{Eq24}) the gaps of system around $\epsilon=\pm\pi/T$ quasi-energies are always closed for the given parameter preventing topological phase transition. For the parameter $\alpha=1$, $\beta=0$, and $\phi=\pi$, the time-glide symmetry is preserved and there are no zero-energy gap as well as zero-energy edge states, see Fig. \ref{fig5}(d). As a result, one can switch between both the edge states alternatively via $\phi$. This interestingly implies that zero-energy or $\pi$-mode edge states originated from, respectively, the static or dynamic parts of the model can be eliminated in a controlled manner.  \begin{figure*}[t!]
    \centering
    \includegraphics[width=.8\linewidth]{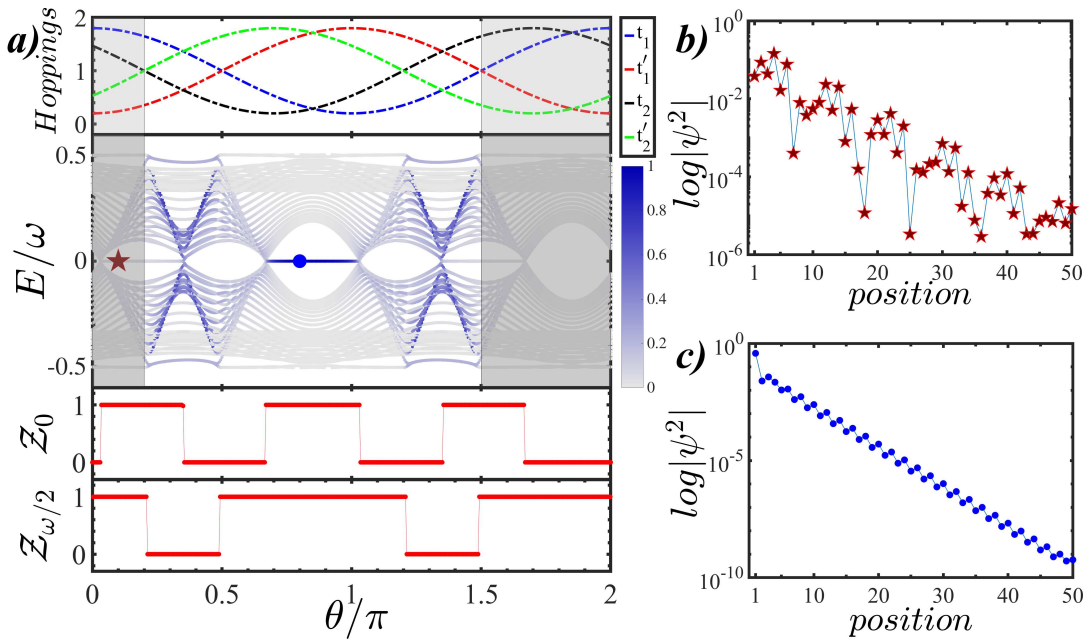}
    \caption{(Color online) (a) Energy spectra and relevant topological invariants as a function of $\theta/\pi$ for $\alpha=\beta=1$, $\phi=0.3\pi$, $T=1.66\pi$, and $t_p=0.5T$ in the first Floquet-BZ. The color scale in the central panel of (a) encodes the weight $\mathcal{W}_0$ on the $n=0$ replica. The upper panel shows the variation of the hoppings in $h_1$ and $h_2$. The dashed gray areas mark the regions where both $h_1$ and $h_2$ are topologically trivial. (b) and (c) show the probability (in log scale) as a function of position associated to the states marked with a star and a dot (respectively) in the central panel of (a).}
    \label{fig8}
\end{figure*}

In Fig. (\ref{fig6}), we examine the influence of shifting the partition time $t_p$ and time periodicity $T$ away from, respectively, $T/2$ and $\pi$ with $\alpha=\beta=1$ and $\phi=\pi$. The top (bottom) panels have plotted for $t_p=0.5T$ ($t_p=0.3T$) with $T=\pi$ for the left column and $T=0.8\pi$ for the right column. For $t_p=0.5T$ and $T=\pi$, the time-glide symmetry is preserved and, as shown in Fig. \ref{fig6}(a), there is no gap at $\epsilon=0$ in the band structure. So, there are no topological edge states at such energy. This is because of the gapless bulk states at $k_s=\pi$ and $\epsilon=0$ predicted by Eq. (\ref{Eq22}). Also, according to Eq. (\ref{Eq21}), the other gap closing occurs at $\epsilon=0$ for any value of $\theta$ at $k_s=0$. However, the gap of system at each $k_s$ would be opened by changing $T$ and $t_p$. In Fig. \ref{fig6}(b), we set $T=0.8\pi$, this causes a gap opening at $k_s=0$. But the presence of time-glide symmetry ensures that the gap at $k_s=\pi$ remains still closed. These result in reducing the density of states around $\epsilon=0$. In Fig. \ref{fig6}(c), for $T=\pi$ and $t_p=0.3T$, there are topological edge states at zero quasi-energy in the range $1.5<\theta/\pi<0.5$ because of breaking time-glide symmetry due to changing $t_p$ from the midpoint. But setting $T=\pi$ being a reason for gap closing at $k_s=0$ leads to the hybridization of the zero-energy edge states with bulk states \cite{FloqTheo14}. As shown in Fig. \ref{fig6}(d), for $T=0.8\pi$ and $t_p=0.3T$, a zero-energy gap opens giving rise the appearance of the localized topological edge states explicitly at zero energy because of simultaneous gap openings both at the supersymmetric points, i.e., $k_s=0$ and $k_s=\pi$. The origin of full gap opening around zero quasi-energy traces back to the time-glide symmetry breaking and the value of $T$ owning to $t_p\ne T/2$ and $T\ne n\pi$, respectively. Note that in Figs. (\ref{fig5}) and (\ref{fig6}), in some cases, the topological invariants for topological edge state at zero or $\pm\pi/T$ quasi-energy show a half integer value stemming from corresponding to the closing of the gap such that their difference for topological and non-topological regimes reveals a quantized value.

As already discussed, $t_p=T/2$ ensures the existence of time-glide symmetry, so the breaking of this symmetry and its consequences are interesting. In Fig. (\ref{fig7}), we have plotted quasi-energy spectra for the case of full and out of phase dimerization during successive time durations, i.e., $\beta=0$ and $\phi=\pi$, with different values of $T$ and $t_p$. The quasi-energy is shown for $t_p<T/2$ and $t_p>T/2$ in the left and right column of Fig. \ref{fig7}, respectively. Also, the driving period for the top (bottom) row is $T=\pi$ ($T=2\pi$). From Figs. \ref{fig7}(a) and \ref{fig7}(b), one can see that topological edge states at zero energy exist for any value of $\theta$ when $t_p<T/2$ without occurring topological phase transitions, while they disappear when $t_p>T/2$. In contrast as presented in Figs. \ref{fig7}(c) and \ref{fig7}(d) with $T=2\pi$, the change of $T$ provides several topological phase transition points at zero and $\pm\pi/T$ quasi-energies. In addition, the ranges of parameters where the zero-energy edge states exist or do not exist can be inverted by changing the time duration $t_p$ from $t_p<T/2$ to $t_p>T/2$. 

\subsection{Generating Floquet topological states from topologically trivial snapshots}

Before concluding this section, we now turn to one more interesting case where a topological states can be generated by topologically trivial snapshots. Figure \ref{fig8}(a) shows the hopping parameters for the dimer defining $h_1$ and $h_2$ (upper panel), the spectrum (central panel), and the invariants (lower panel), i.e., multi-band Zak phase \cite{Zak}, both at the zone center ($\mathcal{Z}_0$) and edge ($\mathcal{Z}_{\omega/2}$), as a function of $\theta$. In the upper and middle panels, the parameter regions where $h_1$ and $h_2$ are both in a trivial regime are marked with gray shaded areas. At $\theta=0$, one starts from a situation where $h_1$ and $h_2$ are topologically trivial. In spite of that, we can see that there are topological edge states forming even at zero energy within the gray shaded areas, in other words, one can generate topological states out of trivial snapshots representing zero-energy Floquet edge states. The associated edge states decay exponentially from the edge, although they have a richer structure of oscillations as compared with those appearing outside of the shaded areas (see the panel of Fig.~\ref{fig8}(b)).

Further scrutiny of the numerical results shows the nature of the topological transitions leading to the Floquet topological edge state in the gray shaded areas. In particular, we find that the first transition point occurring close to $\theta=0$ is due to the closing of the gap between the $n=1$ and $n=-1$ replicas. The second transition point at $\theta\sim 0.35 \pi$, in contrast, is produced by a contribution from the $n=0$ replica (the weights on the $n=0$ replica are shown through the color scale in the central panel in \ref{fig8}(a)). The edge states appearing close to $\theta\sim \pi$ have full weight on the $n=0$ replica and differ from those at $\theta\sim 0.2 \pi$ which have a weight on the $n=\pm 1$ replicas and therefore involve processes of photon emission and absorption. A closer look at the edge reveals is offered in panels (b) and (c) ((b) is for the state marked a star in (a) while (c) corresponds to the circle), revealing a richer structure for the states stemming from the mixing between $n=\pm 1$ replica.
\begin{figure}[t!]
    \centering
    \includegraphics[width=1\linewidth]{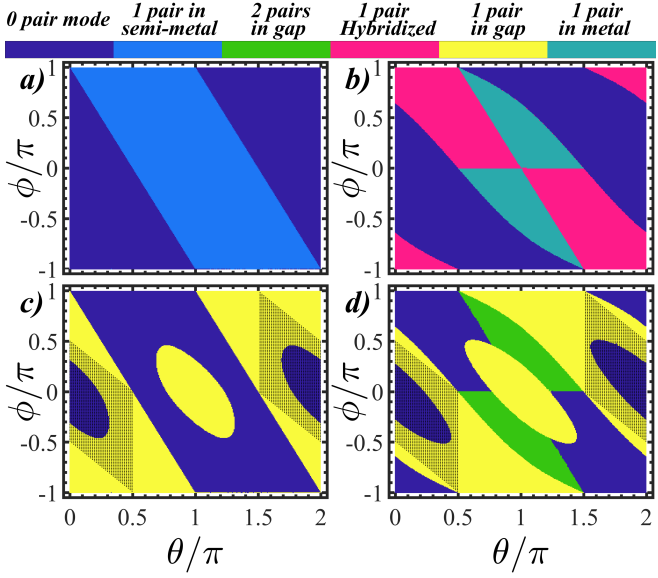}
    \caption{(Color online) Topological phase diagram of the $\epsilon=0$ quasi-energy edge states as functions of ($\theta/\pi, \phi/\pi$) with $\alpha=\beta=1$. For the top (bottom) row $T=\pi$ ($T=1.66\pi$), while for the left (right) column $t_p=T/2$ ($t_p=0.3T$). Dashed black points indicate a region where both the $h_1$ and $h_2$ host a trivial topological phase.}
    \label{fig9}
\end{figure}

\section{Topological phase diagram}\label{s7}

\begin{figure}[t!]
    \centering
    \includegraphics[width=1\linewidth]{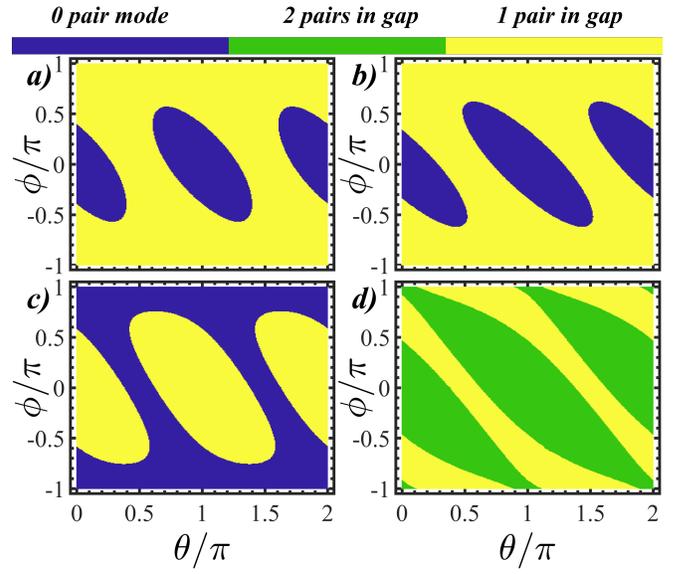}
    \caption{(Color online) Topological phase diagram of the $\epsilon=\pm \pi/T$ quasi-energy edge states as functions of ($\theta/\pi, \phi/\pi$) with $\alpha=\beta=1$. For the top (bottom) row $T=\pi$ ($T=1.66\pi$), while for the left (right) column $t_p=0.5T$ ($t_p=0.3T$).}
    \label{fig10}
\end{figure}

In Figs. \ref{fig9} and \ref{fig10}, we have plotted the phase diagram versus $\theta$ and $ \phi$ for zero and $\pm \pi/T$ energy edge states, respectively, with different values of $T$ of $t_p$ that are characterized by Zak phase.

Figures \ref{fig9}(a) and \ref{fig9}(b) show the phase diagram for $t_p=0.5T$ and $t_p=0.3T$, respectively, with $T=\pi$. The dark and light blue areas show a normal insulator and one pair of topological edge states in the presence  the semimetal phase, respectively. In fact by setting $t_p=T/2$, there is no coupling between the $n=\pm 1$ replicas which touch each other at $k_s=0$ ($T=\pi$). But moving away from the center of the interval, for $t_p=0.3T$ (see Fig. \ref{fig9}(b)) there is a coupling between the $n=\pm 1$. Fig. \ref{fig9}(b) shows the situations where hybridized edge states coexist with bulk states in pink, while the ones where edge states appear in the presence of a metallic state are shown in cyan. Panels (c) and (d) show the phase diagram for $T=1.66\pi$ with the same $t_p$. For $t_p=0.5T$, (see panel (c)) changing $T$ leads to a pair of edge state in the gap of the system which is indicated in yellow. In Fig. \ref{fig9}(d), by changing $t_p$ and $T$ simultaneously, the gap around zero energy opens up thereby preventing the edge states to hybridize with bulk states resulting in two pairs edge states in the gap represented in green color. In addition, the regions where the snapshots for each of the two time durations correspond to trivial topological phases are shown with by black dots.

Finally, Figs. \ref{fig10}(a) and \ref{fig10}(b) show the phase diagram of the $\epsilon=\pm \pi/T$ quasi-energy edge states for $t_p=T/2$ and $t_p=0.3T$, respectively, with $T=\pi$. By changing the partition time $t_p$ away from $T/2$, the range of the trivial region is expanded slightly. Also, the increase of driving period $T$ can make that the nontrivial region becomes more limited (see Fig. \ref{fig10}(c)). In addition, as can be seen from Fig. \ref{fig10}(d), the increase of $T$ and the decrease of $t_p$, simultaneously, can induce two pairs edge states at the $\epsilon=\pm \pi/T$ quasi-energies.

\section {Summary} \label{s8}

In this paper, we studied a periodically quenched dimer where the Hamiltonian switches from $h_1$ to $h_2$ at a partition time $t_p$ during each period $T$. The occurrence of edge states both at zero energy and also at the Floquet-BZ edge (the so called $\pi$ modes) is studied in detail. Our results show that the parameters $t_p$ and $T$ can be used as control parameters resulting in a rich variety of situations including: topological edge states in a gapped system, edge states coexisting with a semimetal due to other replicas, edge states hybridized with a continuum. We have also illustrated how one can obtain topological states out of topologically trivial snapshots. Finally, we discussed the role of the different symmetries in the observed behavior and computed the relevant topological invariants.

\section*{Acknowledgments}
We would like to thank L. Zhou for reading the manuscript and useful comments on this work. LEFFT acknowledges support from FondeCyT (Chile) under grant number 1211038.


\end{document}